\documentclass[12pt]{article}
\usepackage{graphicx}
\begin{document}
\begin{center}
{\huge{Order of precedence and 
age of Y-DNA haplotypes}} \vspace{5mm} \\

{\large
M. Veselsky \vspace{5mm} \\
}

Institute of Physics, Slovak Academy of Science, Bratislava, Slovakia \vspace{5mm} \\

{\bf Abstract}  \\
\end{center}
{
A simple method, inspired by procedures used in physics 
of nuclear multifragmentation, allows to establish order of precedence and 
age of pairs of haplotypes separated by one mutation. 
For both haplotypes of the pair, searches for existing 
haplotypes, differing by increasing number of mutations, 
are carried out using a database. The resulting ratios 
of frequences of haplotypes, found at given mutation distances, 
are compared to calculated probability ratios. The 
order of precedence and age of the pair of haplotypes can be deduced when 
the resulting ratios follow hyperbolic dependence. 
Method can be used with relatively small and not necessarily complete 
samples, using publicly accessible databases. 
}

\section*{Introduction}

The macromolecule DNA is a cornerstone of the life on Earth. 
The part of human DNA, contained in the Y chromosome, does not recombine and 
thus it transfers in the male line unchanged. However, it can mutate 
by spontaneous and irreversible changes in the order of individual 
nucleotides or their sequences, 
and one particular type of mutation leads to change of the number of 
repetitions of specific sections of the DNA in the locations called STR 
markers. 
The present knowledge on the subject implies that such mutation of DNA 
is a stochastic process which can be 
characterized by a rate of mutations per time and thus 
it can be described in analogy to physical phenomena 
such as the radioactive decay. 
Biological processes leading to such mutations are beyond the scope 
of this work and the term haplotype is restricted here to 
a set of numbers which can be changed as a result of mutation, 
according to quantitative laws described by a model introduced below. 

\section*{Model}

The frequence of mutations can be described by 
a constant called mutation rate, which is a direct equivalent of the 
decay rate in the radioactive decay. Mutation rate can be 
defined for each STR marker separately, however it is a common 
practice to define mutation rate for a set of STR markers called 
haplotype. We formulate initial condition that at the start of the 
process there exists $N_0$ copies of a unique haplotype 
within the studied population. In analogy to radioactive decay 
the number of copies of this initial 
haplotype in the population will evolve according to equation 

\begin{equation}
N_0(t) = N_0(t=0) \rm{e}^{ - \lambda t}
\label{eqn0}
\end{equation}

where $t$ is the time and $\lambda$ is the mutation rate ( per 
haplotype ).

Since mutation rate is independent of the number of mutations, 
the number of haplotypes with one mutation $N_1(t)$ can be determined 
using differential equation

\begin{equation}
\frac{dN_1(t)}{dt} + \lambda N_1(t) = \lambda N_0(t)
\label{eqn1}
\end{equation}

which is an 
inhomogeneous linear differential equation and after substituting 
$N_0(t)$ from equation (\ref{eqn0}) one obtains solution 

\begin{equation}
N_1(t) = \lambda N_0(\hbox{t=0}) t \rm{e}^{ - \lambda t}
\label{eqsoln1}
\end{equation}

and in similar way one obtains solution for a 
number of haplotypes with $m$ mutations $N_m(t)$ 

\begin{equation}
N_m(t) = \lambda^{m} N_0(\hbox{t=0}) \frac{t^{m}}{m!} \rm{e}^{ - \lambda t}
\label{eqsolnm}
\end{equation}

for any $m \geq 0$. Obviously a probability can be obtained by dividing 
the equation (\ref{eqsolnm}) by a total number of haplotypes, which remains 
equal to initial number of haplotypes $N_0(\hbox{t=0})$ 

\begin{equation}
P_m(t) = \frac{(\lambda t)^{m}}{m!} \rm{e}^{ - \lambda t} \hbox{ }
\label{eqprobm}
\end{equation}

and one arrives to a Poissonian distribution of mutations at a given 
time $t$ with mean and variance equal to $\lambda t$. Thus in the present 
case the Poissonian distribution is not a statistical approximation but it 
represents analytical solution to the evolution of the system. 

\begin{figure}[h]
\centering
\vspace{5mm}
\includegraphics[width=9.5cm,height=8.75cm]{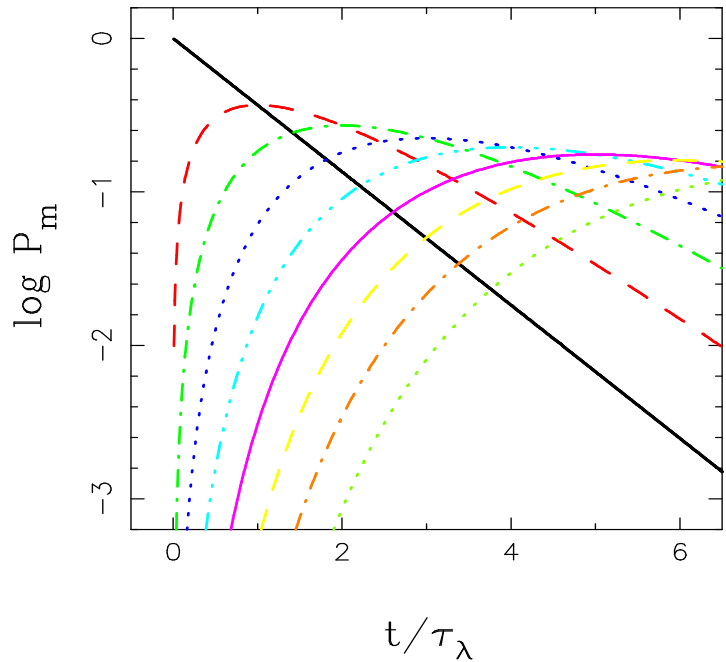}
\caption{
Probabilities of occurence of $m$ mutations 
in the sample at a given time. 
Lines represent probabilities $P_m$ calculated for $m=0 - 8$ 
using equation (\ref{eqprobm}).
}
\label{fpmut}
\end{figure}

The Poissonian probabilities calculated using equation (\ref{eqprobm}) 
are shown in Figure \ref{fpmut} as a family of 
lines representing probabilities of occurence of $m$ mutations 
in the sample at a given time. Time is expressed in units of $t/\tau_{\lambda}$ 
where the average interval between two mutations is obtained as 
$\tau_{\lambda} = 1 / \lambda$. 

One easily recognizes dependence for $m=0$ ( see equation (\ref{eqn0}) ), an 
exponential which is represented by a straight line in the logarithmic scale. 
Other lines follow for increasing values of $m$ and peak correspondingly 
at time $t/\tau_{\lambda} = m$.   

\begin{figure}[h]
\centering
\vspace{5mm}
\includegraphics[width=9.5cm,height=8.75cm]{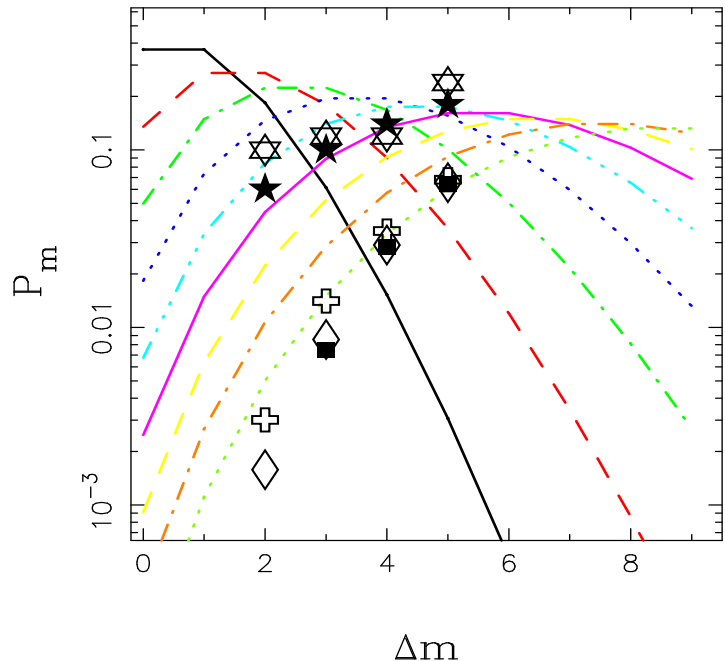}
\caption{
Distribution of mutation distances. Lines - distributions 
calculated using equation (\ref{eqprobm}) for constant time 
step $\tau_{\lambda}$. 
Five- and six-pointed asterisks - haplotypes 
from Slavic countries, distanced from haplotypes $A$ and $B$ 
by $\Delta m$ mutations, found in database Ysearch.org,   
diamonds and squares - all haplotypes, distanced from the haplotype $A$ 
by $\Delta m$ mutations, found in databases Ysearch.org and Ybase.org, 
crosses - all haplotypes, distanced from the haplotype $B$ by $\Delta m$ 
mutations, found in database Ysearch.org. 
}
\label{fdmut}
\end{figure}

Since the most characteristic property of biological systems is their 
fast growth under favorable condition, it is worthwile to note that 
while for simplicity the model was formulated for a population with 
constant total number of haplotypes it can be easily modified 
for exponential growth, even with time-dependent growth rate,
and the resulting Poissonian distribution (\ref{eqprobm}) 
will remain unchanged when the analogue of the 
expression for $N_m(t)$ in equation (\ref{eqsolnm}) will be normalized 
by a total number of haplotypes at a given time. 

\section*{Test with data}

Test was performed with a single Y-DNA haplotype ( originating in Slovakia, 
haplogroup determined as R1b ), which is represented 
by a number of repetitions of nucleotide 
sequences in 17 specific locations on Y-chromosome (STR markers). 
Using this haplotype ( referred as haplotype $A$ ), 
a search for haplotypes distanced by $\Delta m$ mutations   
was performed using the database Ysearch.org \cite{Ysearch}, 
which contains around 
90000 individual records. The distributions of discovered haplotypes 
with specific mutation distances integrated over the 17 STR markers 
are shown in Fig. \ref{fdmut}. 

Lines represent the Poissonian distributions, 
expressed by equation (\ref{eqprobm}), for a given number of mutations 
at times increasing by a constant step of $\tau_{\lambda}$. 
The five-pointed asterisks represent number of occurences of haplotypes 
with a given mutation distance $\Delta m$, originating from the 
Slavic countries, while 
diamonds represent a number of occurences without geographical restrictions, 
dominantly haplotypes from British Isles (mostly from Scotland and
Ireland) and USA. 
Squares represent a number of occurences without geographical restrictions, 
obtained using an alternative database Ybase.org \cite{Ybase}, again 
dominated by haplotypes from British Isles and USA. 

It apears that the results for the Slavic countries 
represent a lower mean number of mutations than the results 
without geographical restrictions, however more detailed analysis 
is difficult due to 
uncertainty in selection of proper normalization of incomplete 
distributions, which further can be convolutions of several components. 
The comparison of results obtained using different databases 
shows consistent agreement, demonstrating reproducibility of the 
procedure. 

For reference, analogous search was performed 
for a haplotype with mutation distance $\Delta m=1$,  
discovered in the database YHRD.org \cite{yhrd} in six 
records distributed in countries with Slavic population  
( further referred as haplotype $B$ ). 
Results of search using the database Ysearch.org 
are shown for the Slavic countries 
and without geographical restrictions, as six-pointed asterisks and crosses, 
respectively. Comparison with the searches, performed using the 
haplotype $A$, indicates shorter mean mutation distance, however it is 
difficult to make unambiguous quantitative conclusions 
without proper normalization.  

\begin{figure}[h]
\centering
\vspace{5mm}
\includegraphics[width=15.5cm,height=8.25cm]{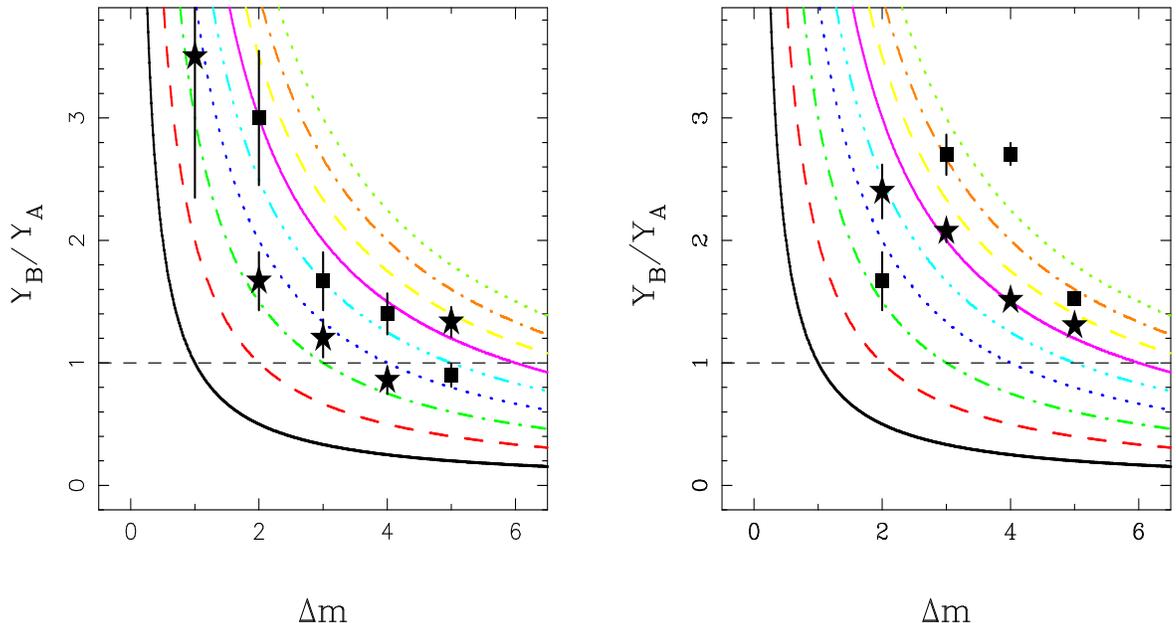}
\caption{
Left panel - Ratios of numbers of haplotypes 
with a given mutation distance
$\Delta m$ for pairs of haplotypes $A$--$B$ (asterisks) 
and  $C$--$A$ (squares) found in the search restricted to Slavic countries, 
lines - calculated probability ratios $P_m$/$P_{m+1}$ for $m=0 - 8$, 
expressed as a function of the mean number of mutations.
Right panel - As in the left panel except that the 
search is without geographical restrictions. 
}
\label{frmut}
\end{figure}

\section*{Probability ratios and hyperbolic scaling}

The situation in Figure \ref{fdmut} illustrates how difficult is to 
make quantitative conclusions without proper 
normalization, due to incomplete mutation distributions 
and possible convolution of several components.  
Situation is quite similar to investigations of nuclear 
reactions, where reconstructed distributions of observables represent 
convolutions of collisions at various impact parameters evolving by different 
reaction mechanisms on various time scales. One possibility to circumvent 
these difficulties 
is to use relative observables such as yield ratios, between 
yields of various reaction products in a given reaction or between 
yields of identical products in two different reactions. An overview of 
methods can be found in \cite{MVIsoTrnd}. 

In the present case, one can attempt to introduce analogous 
procedure. As a starting point, the equation (\ref{eqprobm}) can be used 
to calculate the ratios of Poissonian probabilities for $m+1$ and $m$ 
mutations at a given time $t$ 
 
\begin{equation}
\frac{P_{m+1}(t)}{P_m(t)} = \frac{\lambda t}{m+1}  
= \frac{\frac{t}{\tau_{\lambda}}}{m+1} 
\label{eqrprobm}
\end{equation}

which leads to characteristic hyperbolic time dependences 
documented in Figure \ref{frmut} by a set of lines. 
Such dependences reflect evolution of mutation distance distributions 
in an independent system starting from a single haplotype, 
described by equations (\ref{eqn0}) -- (\ref{eqprobm}), 
for a constant time step $\tau_{\lambda}$. 
In principle one could consider also inverted ratios, leading to 
straight lines with decreasing slopes, nevertheless the representation via  
hyperboles is more sensitive especially for smaller samples 
and mutation distances.  

A corresponding test with data can be performed by calculating ratios of 
mutation distance distributions, shown in Figure \ref{fdmut}, obtained for  
a pair of haplotypes $A$ and $B$, which differ by one mutation. 
The resulting dependences can be compared 
to model values calculated using equation (\ref{eqrprobm}) and 
the difficulties with proper normalization can thus be solved. 
This again reminds procedures often employed in nuclear physics, 
where conditions at specific stage of nuclear multifragmentation, or 
relativistic nucleus-nucleus collisions at the LHC, are reconstructed 
from final observables using the appropriate model assumptions.   
The results of the procedure in the present case are 
shown in Figure \ref{frmut}. In the left panel of Figure \ref{frmut}, 
the asterisks represent the results for haplotypes $A$ and $B$ within  
Slavic countries. 
The squares represent analogous results for a pair of haplotypes, 
where the haplotype $A$ is compared to the only other haplotype with mutation 
distance $\Delta m=1$, found in the database YHRD.org in one record 
(further referred as $C$). 

It is apparent that both dependences assume the expected 
hyperbolic shapes and thus represent the behavior 
described by the model described above. Specifically such 
hyperbolic scaling of the ratios $P_{m+1}(t)/P_m(t)$ means 
that these ratios are consistent with evolution from a single 
haplotype in an independent system. The dependence for the haplotype 
pair $A$--$B$ is located between two 
hyperboles representing the times 3 and 4 $\tau_{\lambda}$, 
except for the last point, which will be discussed later. 
This indicates that within given R1b population 
the two haplotypes are at the initial stage of the haplotype tree 
which evolves independently during a time 3 -- 4 $\tau_{\lambda}$ 
since the moment when the haplotype $A$ emerged or eventually when 
the population carrying this haplotype split from larger R1b population, 
as a result of migration. 
If the order of mutations would be 
opposite, the dependence would assume a corresponding straight line. 
The dependence for the haplotype pair $C$--$A$ 
is located close to the 
hyperbole representing the evolution time 5 $\tau_{\lambda}$ and it 
can be considered as representing the mutation preceding the 
mutation represented by the haplotype pair $A$--$B$. The difference 
of the two "experimental" hyperboles appears by 50 \% larger then 
a mean time gap 
between mutations, however one has to keep in mind that mutation 
is a stochastic process and thus it does not happen at equal time 
intervals. 

\begin{figure}[h]
\centering
\vspace{5mm}
\includegraphics[width=15.5cm,height=8.25cm]{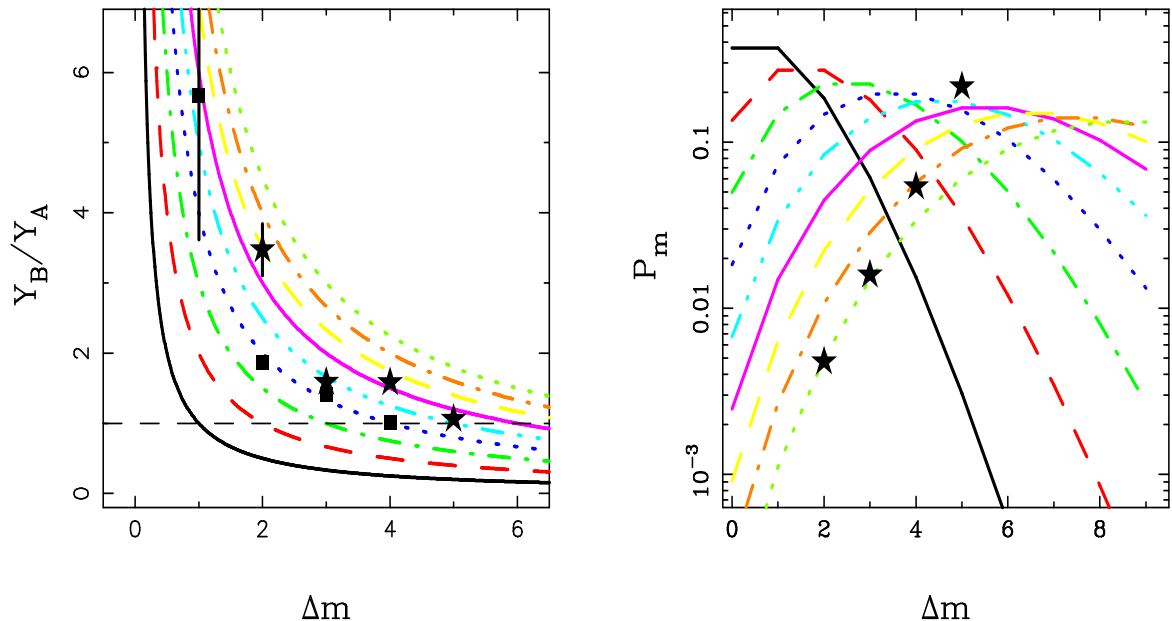}
\caption{
Left panel - Ratios of numbers of haplotypes 
with a given mutation distance
$\Delta m$ for two $\Delta m=1$ pairs of haplotypes from British Isles 
(asterisks and squares) found in the nonrestricted search at Ysearch.org, 
lines - calculated probability ratios $P_m$/$P_{m+1}$ for $m=0 - 8$, 
expressed as a function of the mean number of mutations.
Right panel - Squares - numbers of haplotypes, distanced from haplotype $C$ 
by $\Delta m$ mutations, found at Ysearch.org, 
lines - calculated distributions. 
}
\label{frmutuk}
\end{figure}

On the right panel of the Figure \ref{frmut} asterisks represent again 
the results for the haplotype pair $A$--$B$ while squares 
represent the results for the haplotype pair $C$--$A$, however 
in this case the search is carried out  
without geographical restrictions ( dominated by haplotypes from 
British Isles ). 
Interestingly, in this case the dependence for the haplotype pair $A$--$B$ 
is consistent with hyperbole representing the evolution time 6 $\tau_{\lambda}$ 
and it appears that the common ancestor of haplotypes 
found without geographical restrictions is older from the common 
ancestor of haplotypes found in searches restricted to Slavic countries 
by 2 -- 3 $\tau_{\lambda}$. This can be caused by admixture from 
the descendants of haplotypes closely related to haplotypes $A$--$B$, 
which at mutation distances $\Delta m=3$ start to dominate the trend, 
as documented by the breakdown of 
the hyperbolic dependence without geographic restrictions 
at $\Delta m = 2$ where haplotypes from 
British Isles are not dominant. Thus common ancestor found 
at 6 $\tau_{\lambda}$ is not necessarily the haplotype pair $A$--$B$ 
but some related haplotype, differing by one mutation step early 
in the sequence, which evolved separately, most probably in Western Europe. 

Reciprocally, the dependence for the haplotype pair $A$--$B$ obtained 
without geographical restrictions also appears to explain  
the fact that in the search restricted to Slavic countries 
hyperbolic scaling breaks down at $\Delta m = 5$, since apparently 
at this mutation distance  
the ratio reverts to the trend obtained in the search without 
geographical restrictions shown in the right panel. 

The dependence for the haplotype pair $C$--$A$, obtained 
without geographical restrictions (squares), does not exhibit 
hyperbolic shape, instead it appears to initally grow and 
one can consider it as a mixture of its descendant 
haplotypes and descendants of preceding or contemporary haplotypes, 
as one would expect if the haplotype $C$ was close to the center of haplotype 
distribution at the time of its emergence or separation from the rest 
of population. Thus, the haplotype $C$ appears 
to be a R1b haplotype, transferred by migration or emerging just thereafter 
and admixed into Slavic population, of which the haplotypes 
$A$ and $B$ appear to be subsequent descendants.  

Using the literature \cite{MutRate}, and 
assuming interval of 30 years per generation, 
the value of $\tau_{\lambda}$ can be estimated to approximately 
830 years so the time of independent evolution of the sequence 
of R1b haplotypes starting by haplotype $C$ among the predecessors 
of comtemporary Slavs, determined as 5 $\tau_{\lambda}$, can be estimated 
to 4150 years with uncertainty of about 400 years ( 0.5 $\tau_{\lambda}$ ). 
Furthermore, the age of the common ancestor of all considered R1b haplotypes 
within European population, determined as 6 $\tau_{\lambda}$, can be 
estimated as 5000 years, again with uncertainty of about 400 years.

Since the available data are dominated by haplotypes from British Isles, 
one can try further analysis of this sample by selecting pairs 
of subsequent haplotypes from this area. Two such pairs were 
identified among the results of the searches performed using the 
haplotypes $A$--$C$, first one represented by the records N43KH and 
GXD83 and second one by YNGCV and SAHFV. Both pairs are relatively 
less frequent within the European R1b populations and thus can be rather old. 
Left panel of Figure \ref{frmutuk} shows ratios of numbers of haplotypes 
with a given mutation distance $\Delta m$ for these two pairs of haplotypes 
(squares and asterisks, respectively) found in the nonrestricted search 
at Ysearch.org and one can see a similar situation to the right 
panel of Figure \ref{frmut}. The pair represented by asterisks appears to 
represent similar age as the common ancestor found in the search with pair 
of haplotypes $A$ and $B$ (5000 years), while the other pair is younger 
by approximately 2 $\tau_{\lambda}$, which results in age 3300 years. 
Thus there can be some age structure in this population, possibly 
as a result of subsequent waves of migrations into British Isles. 
This possibility can be reflected by the structure of the haplotype 
distribution. A relatively distant haplotype from outside of the contemporary 
haplotype distribution can serve as a probe and possibly reveal such structure. 
Squares in the right panel of Figure \ref{frmutuk} show distribution of 
mutation distances from a rather old haplotype $C$, obtained by 
search at Ysearch.org. Comparison with shapes of calculated distributions 
(lines) apears to demonstrate that the mutation distance distribution may be 
a convolution of at least two components with unequal total weights, 
of which the younger (more distant) one appears to dominate.  
This could also explain irregularity in the hyperbolic dependence 
represented by asterisks in the left panel of Figure \ref{frmutuk} 
since the left part of the dependence is apparently older and possibly further 
from the centre of contemporary haplotype distribution. 

\begin{figure}[h]
\centering
\vspace{5mm}
\includegraphics[width=9.5cm,height=8.75cm]{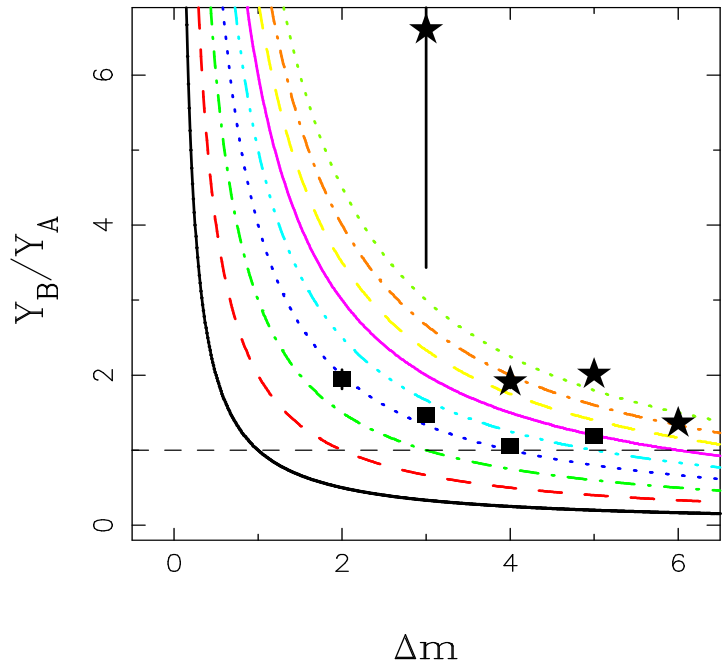}
\caption{
Ratios of numbers of haplotypes with a given mutation distance
$\Delta m$ from two $\Delta m=1$ pairs of R1b haplotypes, 
found in nonrestricted search at Ysearch.org. One haplotype 
of each pair belongs to historical person, Nikolai II Romanov 
and Tutankhamon (squares and asterisks), 
lines - calculated probability ratios $P_m$/$P_{m+1}$ for $m=0 - 9$, 
expressed as a function of the mean number of mutations.
}
\label{frmutkr}
\end{figure}

Based on the above analysis, one can also try to investigate available 
R1b haplotypes, attributed to historical persons. 
In Figure \ref{frmutkr} are shown  
ratios of numbers of haplotypes with a given mutation distance
$\Delta m$ from two $\Delta m=1$ pairs of R1b haplotypes, 
found in nonrestricted search at Ysearch.org. One haplotype 
of each pair belongs to historical person, Nikolai II Romanov 
and Tutankhamon (squares and asterisks). 

The haplotype of Nikolai II Romanov 
(GXK2B) is complemented by a nearest haplotype 
from a person 
of Russian origin, possibly a member of Romanov family (7UFPX), 
and the resulting 
dependence, obtained for this pair in geographically unrestricted search 
( dominated by haplotypes from Western Europe ), 
is close to the one obtained using the 
younger of the two haplotype pairs from Brithish Isles (squares in the right 
panel of Figure \ref{frmutuk}) and thus age of the pair 
of haplotypes can be estimated to 3300 years  
as a time of its existence within population of Western Europe. 
This is indeed  
consistent with German origin of the male line of Romanov family since 18th 
century. 

Since the haplotype of Tutankhamon 
(ER7RQ) is rather distant from the distribution of existing R1b haplotypes 
in the database, it was complemented by a haplotype claimed to belong 
to remains found in archeological location in Lebanon, 
which is mentioned in the supplementary commentary to the record 
in the Ysearch.org database. 
Position on Figure \ref{frmutkr} apparently means that common 
ancestor with the related haplotypes in the distribution of European 
R1b lived before up to 8 -- 9 $\tau_{\lambda}$ ( 6500 -- 7500 years ), 
which might be the time when the ancestors of Tutankhamon split from 
the ancestors of population, which now lives in Europe. 
Obviously, to claim that the haplotype of Tutankhamon is the haplotype 
of the common ancestor would be rather far reaching since data at 
smaller mutation distances are missing and thus accidental later 
closer approach of branches in the mutation tree can not be excluded. 
A larger systematics of R1b haplotypes from Middle East would 
be helpful in further analysis. 

Concerning the method presented here, 
analogy of DNA mutation to radioactive decay and chemical kinetics 
was employed already by A. Klyosov \cite{Klyosov}. 
His method uses equation (\ref{eqn0}) to determine age 
of common ancestor of the whole sample of haplotypes. 
This is achieved by comparison of the total number of haplotypes 
in the sample to the number of mutations within the sample, which is 
obtained by constructing detailed mutation trees for the whole sample. 
In mathematical sense such method relies on integral observables 
while the method presented here uses differential observables 
and can relate individual haplotypes to the bulk 
of the haplotype distribution or its subparts. 
Ultimately both methods are complementary. In his work \cite{Klyosov2}, 
after applying his method separately to R1b populations in various countries, 
Klyosov arrives to conclusion that on its migration to Europe 
the R1b population split once at about 6000 years ago in the 
teritory of Asia Minor while its final expansion into whole 
of Western and Central Europe, and separation into local 
populations occured around 4000 years ago during expansion 
of the archeological Bell Beaker culture. These times of 
migrations, resulting in splitting of R1b population, are in principle 
consistent with the results obtained in this work. 
Both methods thus appear in mutual agreement. 
The method presented in this work is specific by its capability
to perform quick age estimates for individual haplotypes. 
It is suitable specifically for unique haplotypes away 
from the centre of haplotype distributions, 
while for more common haplotypes proper choice of 
the $\Delta m=1$ haplotype will be necessary.  

Klyosov in his work \cite{Klyosov} mentions occurence of reverse mutations, 
which influence the mutation counting and thus distort the final time estimate. 
The model, presented in the present work, relies on comparison of 
observables of subsequent mutation stages and the results can not be 
dramatically influenced by this circumstance. One can still introduce 
a minor correction in the form of reduced mutation rate, 
for a set of 17 STR markers such correction will be less than 2 \% per 
generation.  

In summary, the method, inspired by analogous procedures used in physics 
of nuclear multifragmentation, allows to establish order of precedence and 
determine age of pairs of haplotypes separated by one mutation. 
For both haplotypes of the pair, searches for existing 
haplotypes, differing by increasing number of mutations, 
are carried out using a haplotype database and the resulting ratios 
of frequences of haplotypes, found at given mutation distances, 
are compared to calculated probability ratios. The 
order of precedence and age of the pair of haplotypes can be deduced when 
resulting ratios follow hyperbolic dependence. 
The method provides a simple tool which 
can be used with relatively small and not necessarily complete 
samples, available in publicly accessible databases.

\end{document}